\documentclass[showpacs,twocolumn,showkeys,amsmath,amssymb,pre]{revtex4-1}

\usepackage{bm}
\usepackage{bbm}
\usepackage{color}	
\usepackage{graphicx}   
\newcommand{\rd}{{\mathrm d}}

\newcommand{\ri}{{\mathrm i}}
\newcommand{\muB}{\mu_{\rm B}} 
\newcommand{\kB}{k_{\rm B}}                 

\begin{document}

\title[Floquet-state paramagnetism]
      {Environment-controlled Floquet-state paramagnetism} 

\author{Onno R.\ Diermann$^1$,
	Heinz-J\"urgen Schmidt$^2$,
	J\"urgen Schnack$^3$,
	and Martin Holthaus$^1$}
	
\affiliation{$^1$Carl von Ossietzky Universit\"at, Institut f\"ur Physik,
	D-26111 Oldenburg, Germany}	
\affiliation{$^2$Universit\"at Osnabr\"uck, Fachbereich Physik,
 	D-49069 Osnabr\"uck, Germany}
\affiliation{$^3$Universit\"at Bielefeld, Fakult\"at f\"ur Physik,
	D-33501 Bielefeld, Germany}

\date{May 20, 2020}

\begin{abstract}
We study the response of ideal spin systems which are interacting with both 
a strong oscillating magnetic field, and a thermal environment, to a weak 
probing magnetic field. We demonstrate that even the sign of the resulting 
mean magnetization depends on the amplitude of the driving field, and that its 
absolute value can be significantly larger than the equilibrium magnetization 
in the absence of time-periodic forcing. Since the underlying Floquet-state 
occupation probabilities are determined by the precise form of the system-bath 
coupling, future measurements of such effects have the potential to establish 
a particularly innovative line of research, providing information on 
nonequilibrium thermodynamics, and giving access to quantities which usually 
remain hidden when probing equilibrium systems. 
\end{abstract} 


\maketitle 


\section{Introduction}
\label{S_1}

The subject of quantum thermal magnetism, {\em i.e.\/}, the behavior of 
quantized magnetic moments simultaneously affected by both an external static 
magnetic field and a thermal environment, has been addressed by Brillouin 
already in 1927~\cite{Brillouin27,Pathria11}: The magnetic moment~$\mu$ of 
atoms possessing an electron shell with total angular momentum~$J$ and Land\'e 
$g$-factor $g_J$ takes the form $\mu = g_J \muB J$, where $\muB$ is the Bohr 
magneton. When such magnetic moments are exposed to a homogeneous, constant 
magnetic field of strength~$B_0$, quantization of angular momentum gives rise 
to the energy levels
\begin{equation}
	E_m = -m g_J \muB B_0 \; ,
\label{eq:QAM}	
\end{equation}
with $m = -J, \ldots, J$ denoting the magnetic quantum number. Assuming that
the atoms are interacting with a surrounding of temperature~$T$, these levels 
are occupied according to the universal Boltzmann distribution
\begin{equation}
	p_m^{\rm eq.} = 
	\frac{1}{Z_0}\exp\!\left(-\frac{E_m}{\kB T} \right) \; ,
\label{eq:UBD}
\end{equation}
where $\kB$ is Boltzmann's constant, normalization is ensured by the familiar
canonical partition function $Z_0 = \sum_{m=-J}^J \exp(-E_m/\kB T)$, and the 
superscript ``${\rm eq.}$'' indicates thermal equilibrium. With $N$ magnetic 
moments in a volume~$V$ the resulting magnetization reads
\begin{equation}
	M = \frac{N}{V} g_J \muB \langle m \rangle_{\rm eq.} \; ,
\label{eq:MAG}	
\end{equation}		
where the thermal expectation value
\begin{equation}
	\langle m \rangle_{\rm eq.} = \sum_{m=-J}^J m \, p_m^{\rm eq.}
\label{eq:TEV}	
\end{equation}
can be expressed explicitly in terms of the so-called Brillouin function of 
order~$J$~\cite{Pathria11}. An experimental study of such spin paramagnetism 
with Cr$^{+++}$, Fe$^{+++}$ and Gd$^{+++}$ by Henry in 1952 has resulted in
spectacular agreement with this theoretical prediction~\cite{Henry52}. Thus, 
when accepting equilibrium thermodynamics as a firmly established proposition,
measurement of the thermal magnetization of paramagnetic substances provides 
a striking experimental proof for the space quantization of magnetic 
dipoles~\cite{Henry52}. 

The present deliberations are meant to demonstrate that this time-honored 
reasoning can be logically reversed by novel experiments involving 
{\em oscillating\/} fields which may be so strong that they fall outside 
the regime of linear response, and therefore are {\em not\/} covered by the 
well-established concept of AC susceptibility~\cite{Balanda13,KlikEtAl18,
ToppingBlundell19}:  Accepting the quantization of angular momentum as given 
fact, measurement of the mean magnetization of paramagnetic materials in 
strong time-periodic magnetic fields provides information on a particular 
form of nonequilibrium thermodynamics that is becoming known as ``periodic 
thermodynamics''~\cite{Kohn01}. Such periodic thermo\-dynamics refers to 
quantum systems that are both driven by a time-periodic external force, 
and interacting with a heat bath of fixed temperature~\cite{GrahamHuebner94,
BreuerPetruccione97,HoneEtAl09,KetzmerickWustmann10,GasparinettiEtAl13,
LangemeyerHolthaus14,BulnesCuetaraEtAl15,ShiraiEtAl15,RestrepoEtAl16,
HartmannEtAl17,ZhangEtAl17,ChoiEtAl17,Schmidt20,IkedaSato20}. As a consequence 
of time-periodic forcing, the driven system possesses a complete set of 
Floquet states; as a consequence of their interaction with the heat bath, 
a nonequilibrium steady state establishes itself which is characterized by a 
quasi\-stationary distribution of Floquet-state occupation probabilities. 
Quite unlike the universal Boltzmann distribution~(\ref{eq:UBD}), such 
Floquet-state distributions depend on the very details of the system-bath 
interaction~\cite{Kohn01,BreuerEtAl00,DiermannEtAl19,DiermannHolthaus19}. 
Therefore, measurements of these distributions, or of observables directly 
governed by them, provide experimental access to system-bath coupling 
mechanisms, or to details of the heat bath itself, which remain unobservable 
when probing equilibrium systems. Here we outline the ramifications of 
this general concept for driven spin systems, extending our previous 
work~\cite{SchmidtEtAl19}. We start with a brief sketch of the theoretical 
basics in Sec.~\ref{S_2}, then discuss numerical model calculations providing 
cases in point in Sec.~\ref{S_3}, and summarize the key issues in 
Sec.~\ref{S_4}.

\section{The concept of quasithermal magnetism}
\label{S_2}

\begin{figure}[t]
\centering
\includegraphics[width=0.9\linewidth]{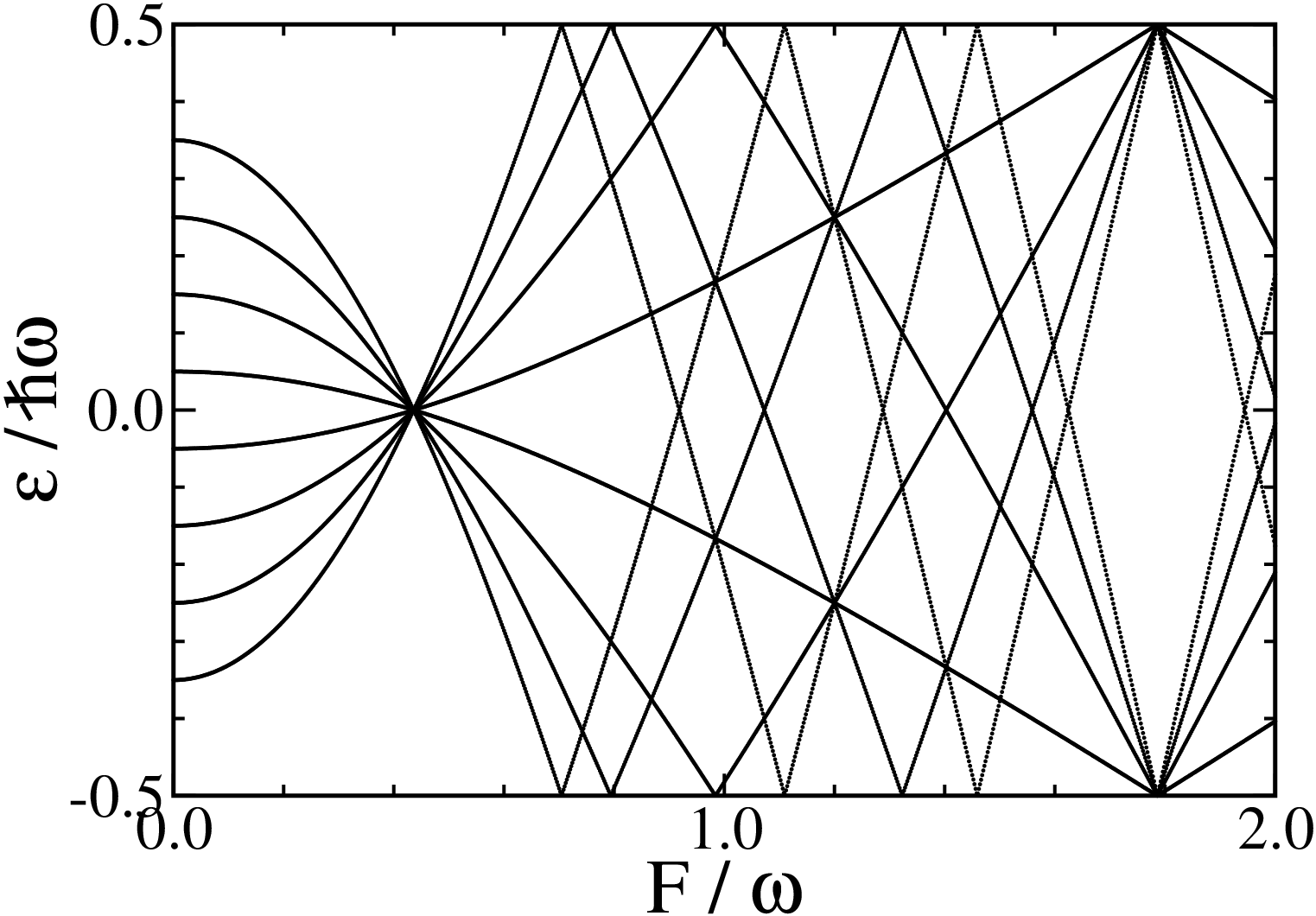}
\caption{One Brillouin zone of quasienergies for a spin $J = 7/2$ driven  
	by a right-circularly polarized high-frequency magnetic field according
	to Eq.~(\ref{eq:CPF}), and exposed to a static magnetic field of scaled 
	strength $\omega_0/\omega = 0.1$. Observe the complete quasienergy
	collapse at $F/\omega \approx 0.44$, indicating the principal resonance 
	$\Omega/\omega = 1$. Further particularly high degeneracies appear at 
	$F/\omega = 1.2$ ($\Omega/\omega = 3/2$) and $F/\omega \approx 1.79$ 
	($\Omega/\omega = 2$).}   
\label{F_1}
\end{figure}

We write the Hamiltonian of a spin~$J$ acted on by both a static and an
oscillating magnetic field in the form          
\begin{equation}
	H(t) = H_0 + H_{\rm osc}(t) \; ,
\label{eq:HAM}
\end{equation}
where 
\begin{equation}
	H_0 = \hbar\omega_0 S_z \; ,
\end{equation}
with $S_x$, $S_y$, $S_z$ denoting the components of the dimensionless spin 
operator, so that the frequency $\omega_0 = -g_J \muB B_0/\hbar$ encodes the 
strength of the static field~$B_0$ oriented along the $z$-axis, in accordance 
with Eq.~(\ref{eq:QAM}). Note that $\omega_0$ may adopt both signs, depending 
on the sign of $g_J$. Considering, to begin with, an oscillating field which is 
right-circularly polarized in the $x$-$y$-plane with angular frequency~$\omega$
and amplitude $B_1$, one has
\begin{equation}
	H_{\rm osc}^{(r)}(t) = 
	\hbar F \big( S_x \cos(\omega t) + S_y \sin(\omega t) \big) \; ,
\label{eq:CPF}
\end{equation}
thus introducing a further frequency $F = g_J \muB B_1/\hbar$. The Floquet
states 
\begin{equation}
	| \psi_m(t) \rangle = 
	|u_m(t) \rangle \exp(-\ri\varepsilon_m t/\hbar)
\end{equation}
of this system, that is, the complete set of solutions of the time-dependent 
Schr\"odinger equation pertaining to the Hamiltonian~(\ref{eq:HAM}) with 
periodic Floquet functions $|u_m(t) \rangle = |u_m(t + 2\pi/\omega) \rangle$, 
have been determined analytically~\cite{SchmidtEtAl19}. Their quasienergies 
are given by
\begin{equation}
	\varepsilon_m = \left[ \frac{\hbar\omega}{2}\right]
	\pm m \hbar\Omega 
	\qquad \bmod \hbar\omega \; ,  
\label{eq:QEC}
\end{equation}  
where the contribution $\hbar\omega/2$ in square brackets figures for 
half-integer~$J$ only and vanishes otherwise;
\begin{equation}
	\Omega = \sqrt{(\omega_0 - \omega)^2 + F^2}
\end{equation}
is the Rabi frequency, and the $\pm$-sign may conventionally be selected 
such that $\varepsilon_m$ reduces to the energy~$E_m$ (modulo $\hbar\omega$) 
when $F$ goes to zero. Figure~\ref{F_1} depicts an interval
$-\hbar\omega/2 \leq \varepsilon < + \hbar\omega/2$ of this quasienergy 
spectrum~(\ref{eq:QEC}), corresponding to one Brillouin zone, for $J = 7/2$ 
and $\omega_0/\omega = 0.1$ {\em vs.\/} the scaled driving strength $F/\omega$. 
In this case of high-frequency driving with positive~$\omega_0$, that is, for 
$\omega \gg \omega_0 > 0$, the canonical representatives of the quasienergy 
eigenvalues, {\em i.e.\/}, those representatives which connect continuously 
to the energy eigenvalues of~$H_0$, effectively {\em attract\/} each other 
when the driving amplitude is increased from zero to small finite values, 
and cross. This complete quasienergy collapse indicates the principal 
resonance $\Omega = \omega$, implying 
$(F/\omega)^2 = 2\omega_0/\omega - (\omega_0/\omega)^2$, so that all 
quasienergies are degenerate (modulo $\hbar\omega$). Also visible in 
Fig.~\ref{F_1} is another conspicuous resonance for which $\Omega = 2\omega$,
and a further, less obvious resonance occuring when $\Omega  = 3\omega/2$, 
giving rise to two points of degeneracy in the Brillouin zone, separated by
$\hbar\omega/2$.

The Hamiltonian~(\ref{eq:HAM}) of the driven spin acts on a Hilbert space
${\mathcal H}_{\rm sys}$ of finite dimension $2J+1$. We now couple this 
system weakly to a heat bath consisting of infinitely many, thermally occupied 
harmonic oscillators, described by a bath Hamiltonian $H_{\rm bath}$ on a
corresponding Hilbert space ${\mathcal H}_{\rm bath}$. Let us briefly summarize
the required formal steps~\cite{BreuerEtAl00,DiermannEtAl19,SchmidtEtAl19}:
The total Hamiltonian on the composite space 
${\mathcal H}_{\rm sys} \otimes {\mathcal H}_{\rm bath}$    
can be written as~\cite{BreuerPetruccione02}
\begin{equation}
	H_{\rm total}(t) = H(t) \otimes \mathbbm{1} 
	+ \mathbbm{1} \otimes H_{\rm bath} + H_{\rm int} \; . 
\end{equation}
For the sake of definiteness, here we choose a bilinear system-bath coupling 
of the form  
\begin{equation}
	H_{\rm int} = V \otimes \sum_{\widetilde\omega} 
	\left( b_{\widetilde\omega}^{\phantom\dagger} 
	+ b_{\widetilde\omega}^\dagger \right) \; ,
\label{eq:SBC}
\end{equation}	 
where 
\begin{equation}
	V = \gamma_x S_x + \gamma_y S_y + \gamma_z S_z
\end{equation}	
with adjustable coefficients $\gamma_x$, $\gamma_y$, $\gamma_z$ carrying the 
dimension of energy, and where $b_{\widetilde\omega}^{\phantom\dagger}$ 
($b_{\widetilde\omega}^\dagger$) denotes an annihilation (creation) operator 
for a bath oscillator with frequency~$\widetilde\omega$; the sum in 
Eq.~(\ref{eq:SBC}) extends over all such oscillators. 
Naturally, the constraint that the system-bath coupling be weak requires
$|\gamma_{x,y,z}| \ll \hbar|\omega_0|$.

Writing the Fourier decomposition of a transition matrix element between
Floquet states $i$ and $f$ as
\begin{equation}
 	\langle u_f(t) | V | u_i(t) \rangle
	= \sum_{\ell\in{\mathbbm Z}} \, V_{fi}^{(\ell)}
	\exp(\ri\ell\omega t) \; ,
\end{equation}
and introducing the corresponding transition frequencies
\begin{equation}
	\omega_{fi}^{(\ell)} = (\varepsilon_f - \varepsilon_i)/\hbar 
	+ \ell\omega \; ,
\label{eq:FTF}		
\end{equation}
one finds the partial transition rates
\begin{equation}
	\Gamma_{fi}^{(\ell)} = \frac{2\pi}{\hbar^2} \, | V_{fi}^{(\ell)} |^2 \,
	N(\omega_{fi}^{(\ell)})  \, \varrho(|\omega_{fi}^{(\ell)}|) \; . 	
\label{eq:PAR}
\end{equation}
Here the thermal expectation values
\begin{equation}
	N(\widetilde\omega) = 
	\frac{\pm 1}{\exp(\beta\hbar\widetilde\omega) - 1}
\label{eq:MON}	
\end{equation}
quantify the mean occupation numbers of the bath oscillators, with 
$\beta = 1/(\kB T)$ indicating the inverse bath temperature. The plus (minus) 
sign applies to positive (negative) transition frequencies~$\widetilde\omega$; 
observe that each transition among Floquet states is associated with an 
infinite ladder~(\ref{eq:FTF}) of such frequencies. Moreover, the 
quantity~$\varrho(\widetilde\omega)$ entering the partial rates~(\ref{eq:PAR}) 
denotes the spectral density of the bath oscillators. The total rate of 
transitions among the Floquet states $i$ and $f$ then is given by the sum 
\begin{equation}
	\Gamma_{fi} = \sum_{\ell\in{\mathbbm Z}} \, \Gamma_{fi}^{(\ell)}	
	\; .
\end{equation}
The desired quasistationary distribution $\{ p_n \}$ of Floquet-state 
occupation probabilities which results from the combined effects of 
time-periodic driving and bath-induced transitions, and which in periodic
thermodynamics replaces the  equilibrium Boltzmann distribution~(\ref{eq:UBD}), 
finally is obtained as a solution to the master equation
\begin{equation}
	\sum_m \big( \Gamma_{nm} p_m - \Gamma_{mn} p_n \big) = 0 \; .	
\end{equation}	
The golden rule for Floquet states, which underlies the partial 
rates~(\ref{eq:PAR}), requires that the bath is not affected by the 
time-periodic driving field~\cite{LangemeyerHolthaus14}, in the sense of the 
usual Born approximation. If this condition is not met, and the bath itself 
``feels'' the drive, one may obtain different results, as has been exemplified 
in Ref.~\cite{GrahamHuebner94}.
The analytical calculation of such Floquet-state occupation probabilities
is feasible for certain integrable systems only~\cite{LangemeyerHolthaus14,
Schmidt20,BreuerEtAl00,DiermannEtAl19,SchmidtEtAl19}; in general, the above 
framework has to be implemented numerically.

Thus, when an ideal paramagnetic material is strongly driven by an oscillating
field $B_1$, while interacting with a thermal environment and being probed by 
a static field $B_0$ directed along the $z$-axis, the resulting nonequilibrium 
``quasithermal'' magnetization is still given by a relation of the general
form~(\ref{eq:MAG}), but the usual thermal expectation value~(\ref{eq:TEV}) 
now has to be replaced by the Floquet-state expectation 
value~\cite{SchmidtEtAl19}
\begin{equation}
	\langle \! \langle m \rangle \! \rangle = \sum_{m=-J}^J
	\langle \! \langle u_m(t) | S_z | u_m(t) \rangle \! \rangle \; p_m
	\; ,
\label{eq:QTA}
\end{equation}
where double brackets indicate one-cycle averaging,
\begin{equation}
	\langle \! \langle u_m(t) | S_z | u_m(t) \rangle \! \rangle
	= \frac{\omega}{2\pi} \int_0^{2\pi/\omega} \!\!\!\!\!\! 
	\rd t \; \langle u_m(t) | S_z | u_m(t) \rangle \; .
\label{eq:TAV}
\end{equation}	
As emphasized above, the particular interest in this scenario stems from the 
observation that, in contrast to their equilibrium antecessors~(\ref{eq:UBD}), 
the Floquet-state occupation probabilities~$p_m$ are not universal, but do
depend on the way the driven system is interacting with its environment. 
We suggest that this unfamiliar, but experimentally detectable fact may be 
exploited in a twofold manner: Measuring Floquet-state expectation values, 
such as the quasithermal averages~(\ref{eq:QTA}), provides information on the 
system-bath coupling, or on properties of the bath. On the other hand, 
deliberate engineering of the system-bath coupling may allow one to create 
systems exhibiting unusual forms of magnetic response.

\section{Numerical examples}
\label{S_3}

\begin{figure}[t]
\centering
\includegraphics[width=0.9\linewidth]{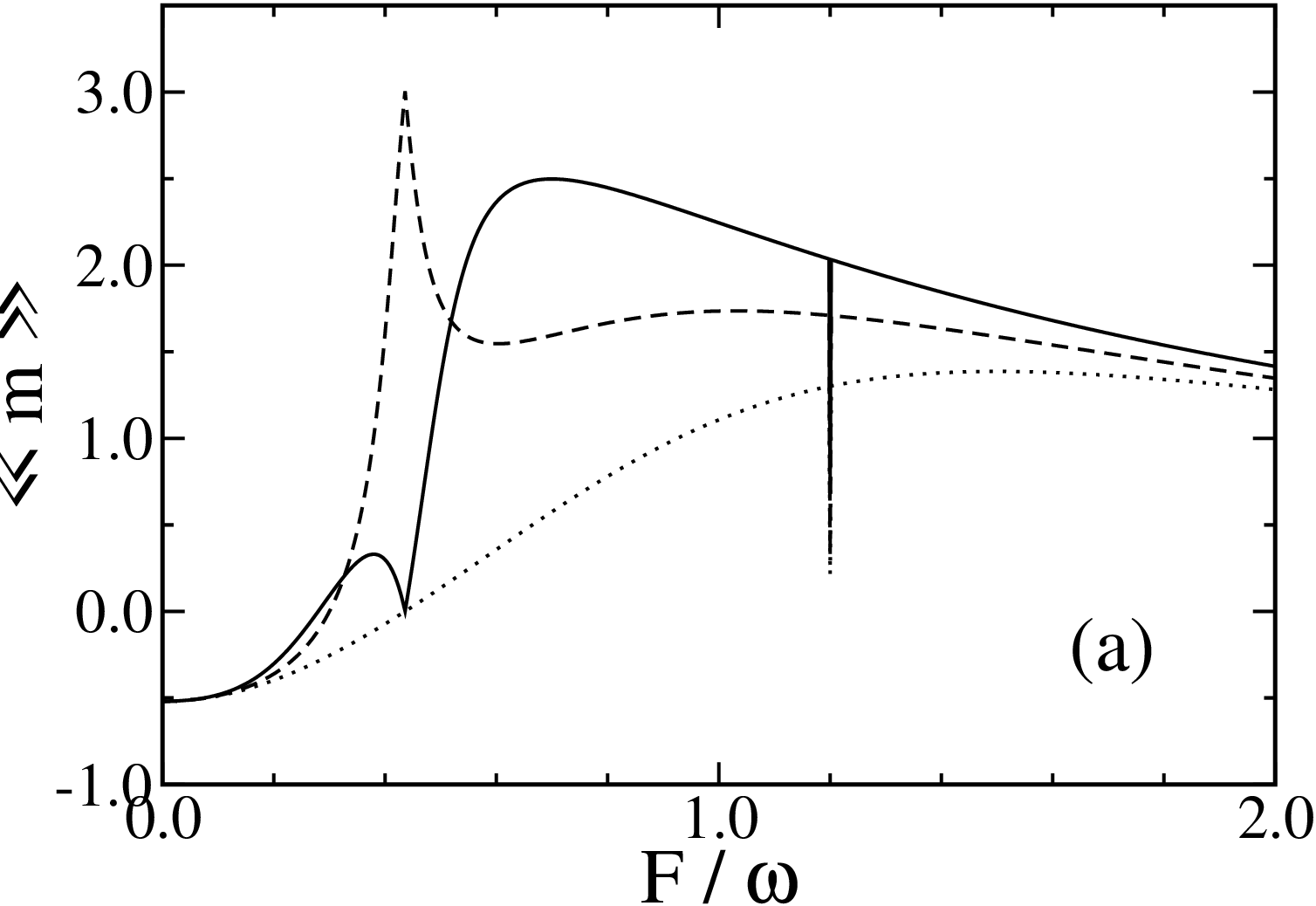}

\vspace{7ex}

\includegraphics[width=0.9\linewidth]{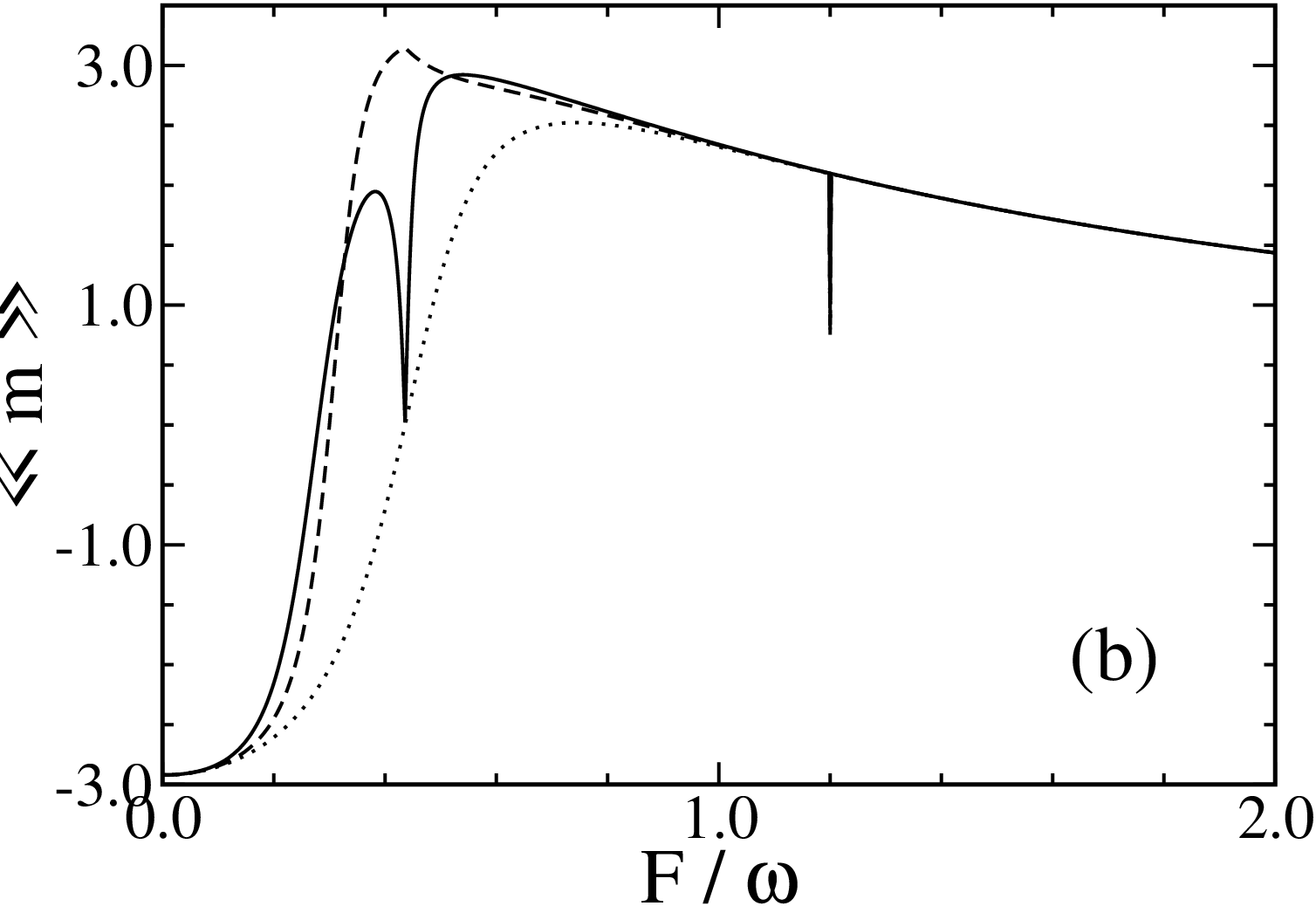}
\caption{Quasithermal expectation values~(\ref{eq:QTA}) for $J = 7/2$,
	$\omega_0/\omega = 0.1$, and right-circularly polarized driving 
	described by Eq.~(\ref{eq:CPF}). Here $\gamma_y = \gamma_z = 0$; the 
	scaled bath temperature is $\kB T/(\hbar\omega) = 1.0$ in the upper 
	panel~(a), and $\kB T/(\hbar\omega) = 0.1$ in the lower one~(b). The 
	bath density of states is taken to be constant (dotted), quadratic 
	(dashed) according to Eq.~(\ref{eq:SOD}), and Gaussian (full lines) 
	with $\omega_{\rm c}/\omega = 5.0$, see Eq.~(\ref{eq:GAD}).}  
\label{F_2}
\end{figure}

We now illustrate the above general concept with the help of selected model 
calculations, intending to highlight typical novel effects. In Fig.~\ref{F_2} 
we depict quasithermal expectation values~(\ref{eq:QTA}) for $J = 7/2$,  
$\omega_0/\omega = 0.1$, and right-circularly polarized driving~(\ref{eq:CPF}),
as corresponding to the spectrum shown in Fig.~\ref{F_1}. In this case the 
Floquet matrix elements $\langle u_m(t) | S_z | u_m(t) \rangle$ do not 
depend on time~\cite{SchmidtEtAl19}, eliminating the necessity to compute 
their temporal averages~(\ref{eq:TAV}). Here we have set, somewhat 
arbitrarily, $\gamma_y = \gamma_z = 0$, and consider three spectral bath 
densities as three prototypical examples for different environments: 
A constant density $\varrho(\widetilde\omega) = \varrho_0$ (dotted lines), 
a superohmic quadratic density   
\begin{equation}
	\varrho(\widetilde\omega) = \varrho_0 
	\left(\widetilde\omega/\omega\right)^2
\label{eq:SOD}	
\end{equation}
(dashed lines), and a Gaussian density
\begin{equation}
	\varrho(\widetilde\omega) = \varrho_0
	\exp\big(-(\widetilde\omega - \omega_{\rm c})^2/2\omega^2\big)	
\label{eq:GAD}
\end{equation}	
with center frequency $\omega_{\rm c}/\omega = 5.0$ (full lines). For both
bath temperatures presupposed in Fig.~\ref{F_2} one encounters a striking 
manifestation of the non-universality of periodic thermodynamics: While all 
three densities necessarily lead to the same magnetization~(\ref{eq:TEV}) in 
ordinary thermodynamics, {\em i.e.\/}, for vanishing driving amplitude, namely, 
$\langle m \rangle_{\rm eq} \approx -0.52$ for $\kB T/(\hbar\omega) = 1.0$ and 
$\langle m \rangle_{\rm eq} \approx -2.92$ for $\kB T/(\hbar\omega) = 0.1$,   
the quasithermal expectation values differ distinctly from each other when 
the drive is turned on. Since positive $\omega_0$ correspond to negative 
$g$-factors, the spin tends to align antiparallelly to the applied static field
when the oscillating field vanishes, requiring negative expectation values 
$\langle\!\langle m \rangle\!\rangle$ for $F/\omega \to 0$. Remarkably, in all
three cases the sign of the magnetization changes when the oscillating field 
becomes sufficiently strong~\cite{SchmidtEtAl19}, as if the sign of $g_J$ 
were reversed through the application of a strong time-periodic drive. The
most notable differences enforced by the different environments appear close to
the principal resonance, $\Omega/\omega = 1$. While the response of a system
with constant density of states to the probing field is not particularly 
affected by this resonance, but vanishes when $\Omega = \omega$, the response 
of a system with Gaussian density of states exhibits a pronounced dip at this 
resonance, again vanishing exactly for $\Omega = \omega$. In marked contrast, 
the superohmic quadratic density of states gives rise to a maximum of the 
effective mean magnetization when $\Omega = \omega$. These differences can 
be explained by analyzing the individual factors $|V_{fi}^{(\ell)}|^2$, 
$N(\omega_{fi}^{(\ell)})$, and $\varrho(|\omega_{fi}^{(\ell)}|)$, which easily 
conspire to affect the partial rates~(\ref{eq:PAR}) in a non-obvious manner, 
with the density of states serving to enhance or suppress these partial rates 
for certain~$\ell$~\cite{DiermannHolthaus19}. The resonance at 
$\Omega/\omega = 3/2$ only reveals itself through sharp spikes in this 
Fig.~\ref{F_2}, whereas no signal can be seen for $\Omega/\omega = 2$.
We remark that the golden rule-based approach leading to the partial 
rates~(\ref{eq:PAR}) might seem endangered close to a resonance, since a 
vanishing transition frequency~(\ref{eq:FTF}) entails a diverging expectation
value~(\ref{eq:MON}). However, the Floquet-state occupation probabilities 
then are determined by ratios of rates. According to the detailed analysis 
performed in Ref.~\cite{Schmidt20} these ratios actually remain finite despite 
the divergence of both numerator and denominator, strongly suggesting that the
golden rule does not lose its validity on resonance.  

From an experiment-oriented viewpoint it deserves to be pointed out that 
the maximum absolute value of the quasithermal magnetization deduced from 
Fig.~\ref{F_2} for comparatively high bath temperature, 
$\kB T = \hbar\omega = 10 \, \hbar\omega_0$, is comparable to the bare 
equilibrium magnetization occurring at much lower temperature,
$\kB T =  \hbar\omega/10 = \hbar\omega_0$. Hence, application of an 
oscillating field can strongly enhance the magnetization of a paramagnet, 
as corresponding to an effective cooling of the system.

\begin{figure}[t]
\centering
\includegraphics[width=0.9\linewidth]{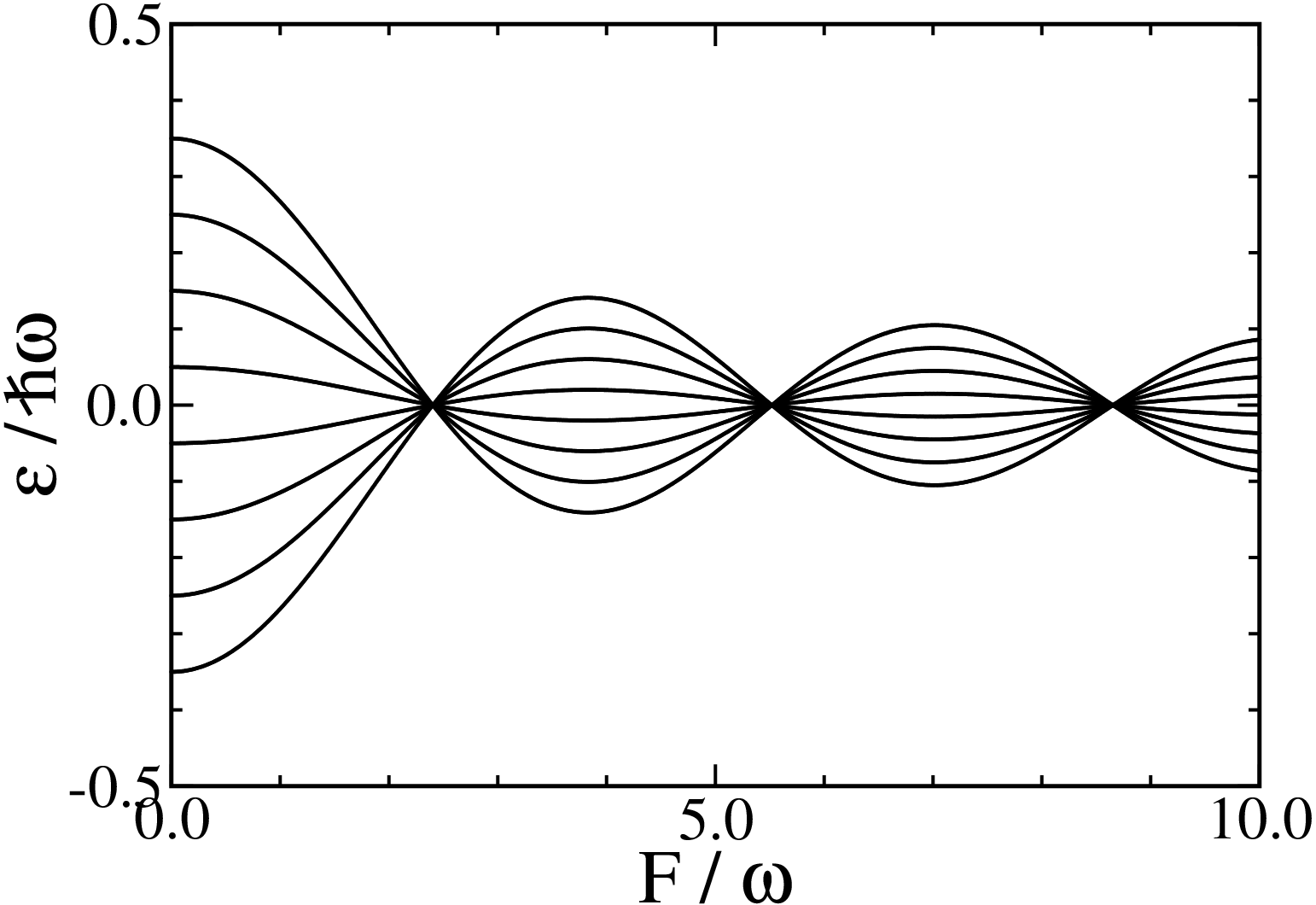}
\caption{One Brillouin zone of quasienergies for a linearly driven spin 
	$J = 7/2$ with $\omega_0/\omega = 0.1$. Observe how the scale 
	of the abscissa here differs from that in Fig.~\ref{F_1}.}   
\label{F_3}
\end{figure}

The sign change of the quasithermal magnetization is related to the crossing 
of all quasienergies observed in Fig.~\ref{F_1} at the principal resonance.
In order to further illustrate this connection we also consider a linearly 
polarized driving field applied orthogonally to the static one, 
\begin{equation}
	H_{\rm osc}^{(lin)}(t) = \hbar F S_x \cos(\omega t) \; .
\label{eq:LPF}	
\end{equation}
The corresponding quasienergy spectrum, again for $J = 7/2$ and 
$\omega_0/\omega = 0.1$, is shown in Fig.~\ref{F_3}. This spectrum is 
described to fair precision by the high-frequency approximation
\begin{equation}
	\varepsilon_m = E_m \, {\rm J}_0(F/\omega)  
	\qquad \bmod \hbar\omega   
\end{equation}
known from tight-binding chains with nearest-neighbor coupling under periodic
driving~\cite{Holthaus92}, with ${\rm J}_0(z)$ denoting the Bessel function of 
the first kind of order zero. Thus, here one finds a sequence of complete 
quasienergy level crossings at the zeros of ${\rm J}_0$.

\begin{figure}[t]
\centering
\includegraphics[width=0.9\linewidth]{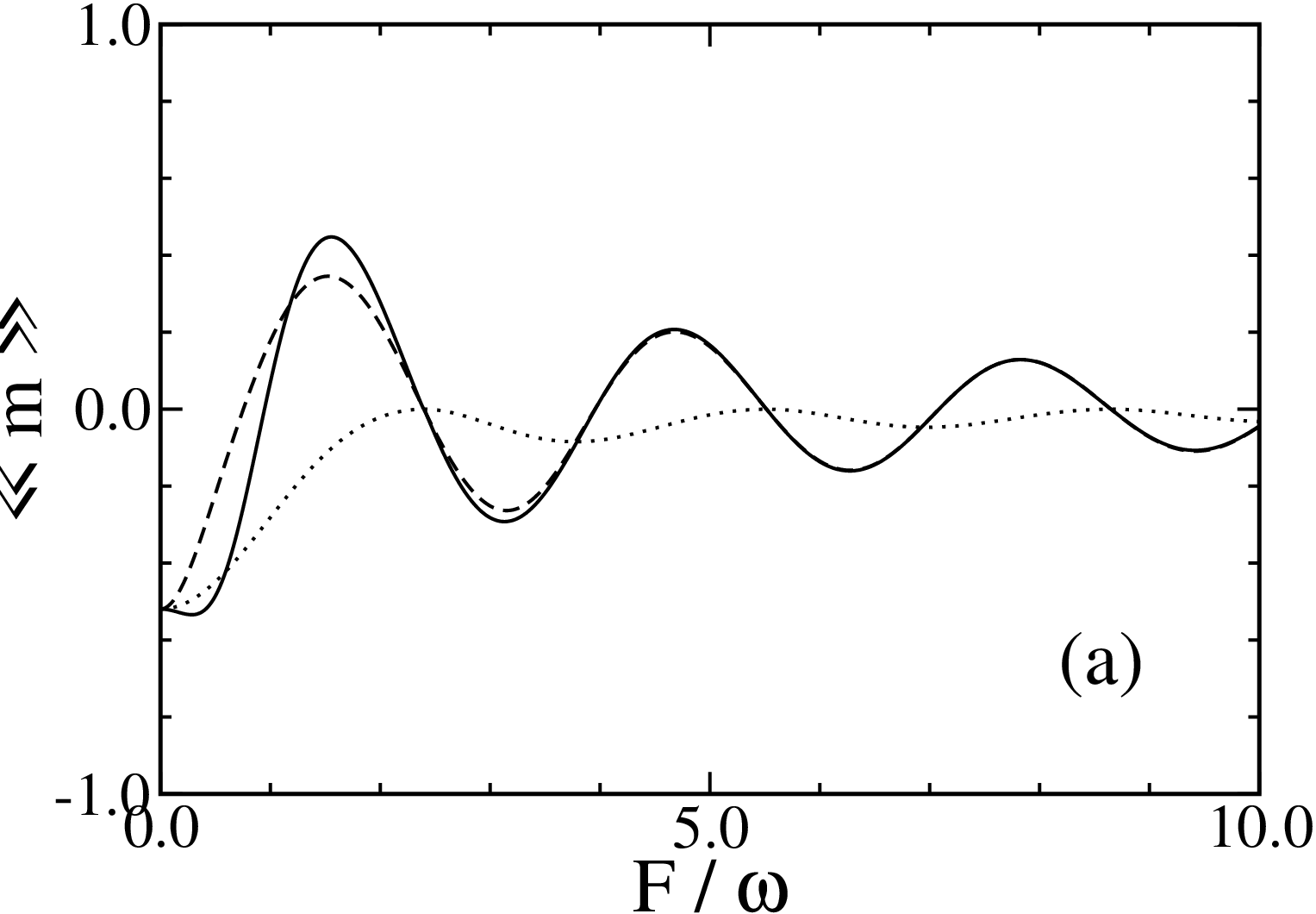}

\vspace{7ex}

\includegraphics[width=0.9\linewidth]{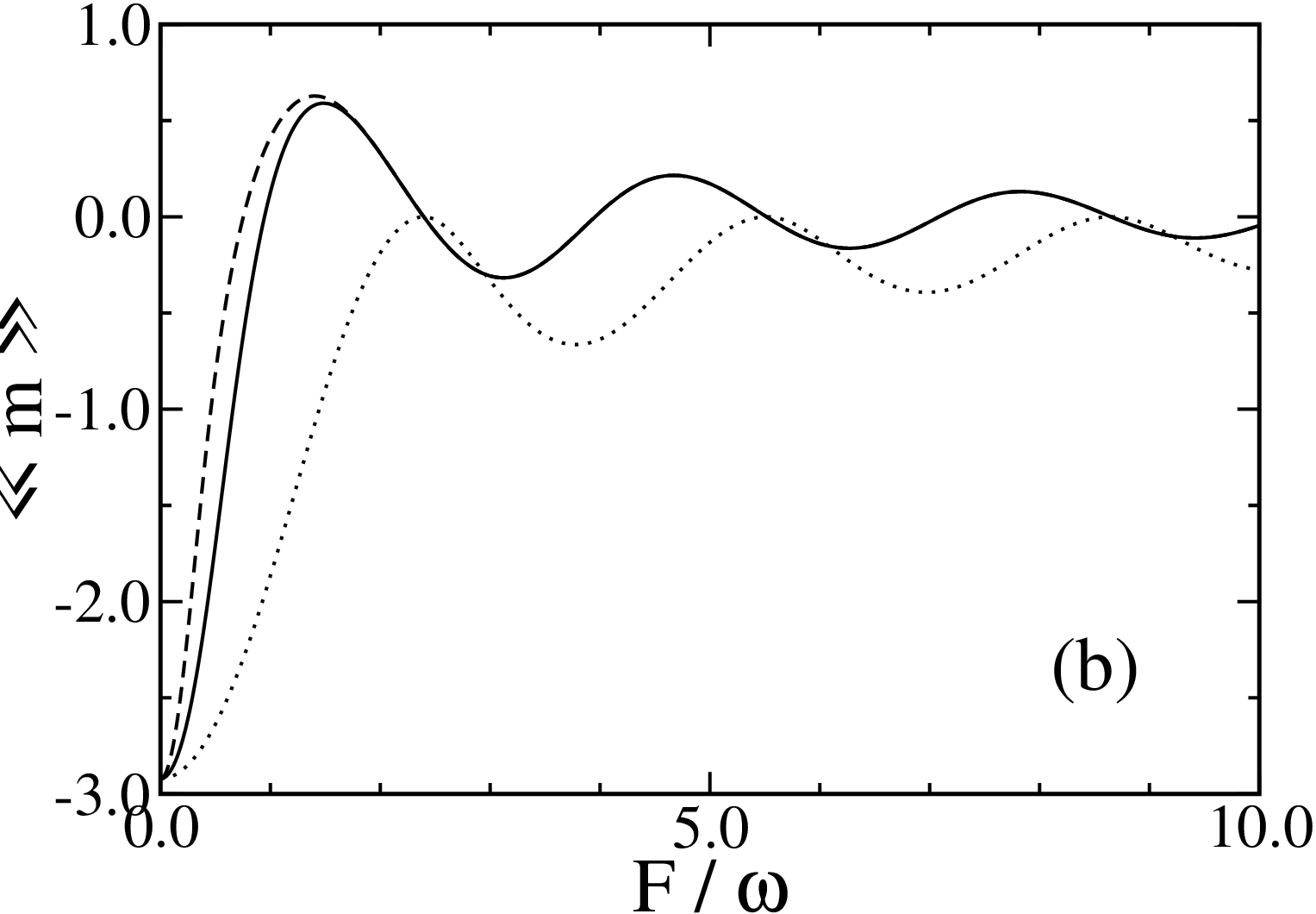}
\caption{Quasithermal expectation values~(\ref{eq:QTA}) for $J = 7/2$,
	$\omega_0/\omega = 0.1$, and linearly polarized driving described by 
	Eq.~(\ref{eq:LPF}). Here $\gamma_x = \gamma_y = \gamma_z$; the scaled
	bath temperature is $\kB T/(\hbar\omega) = 1.0$ in the upper panel~(a), 
	and $\kB T/(\hbar\omega) = 0.1$ in the lower one~(b). The  bath density 
	of states is taken to be constant (dotted), quadratic (dashed) 
	according to Eq.~(\ref{eq:SOD}), and Gaussian (full lines) with 
	$\omega_{\rm c}/\omega = 5.0$, see Eq.~(\ref{eq:GAD}).}  
\label{F_4}
\end{figure}

Now assuming isotropic coupling $\gamma_x = \gamma_y = \gamma_z$, and 
employing the same three bath densities $\varrho(\widetilde\omega)$ as 
before, the resulting quasithermal expectation values~(\ref{eq:QTA}) 
are plotted in Fig.~\ref{F_4}. Interestingly, for both temperatures 
$\kB T/(\hbar\omega) = 1.0$ and $\kB T/(\hbar\omega) = 0.1$ there is no 
sign change of the effective magnetization when the density of states is 
constant, but $\langle\!\langle m \rangle\!\rangle$ becomes zero at the 
quasienergy collapse points observed in Fig.~\ref{F_3}. In contrast, both
the quadratic density~(\ref{eq:SOD}) and the Gaussian density~(\ref{eq:GAD}) 
give rise to repeated sign changes of the magnetization when the driving 
amplitude is increased; note that in any case the quasithermal magnetization 
vanishes at the collapse points.

\begin{figure}[t]
\centering
\includegraphics[width=0.9\linewidth]{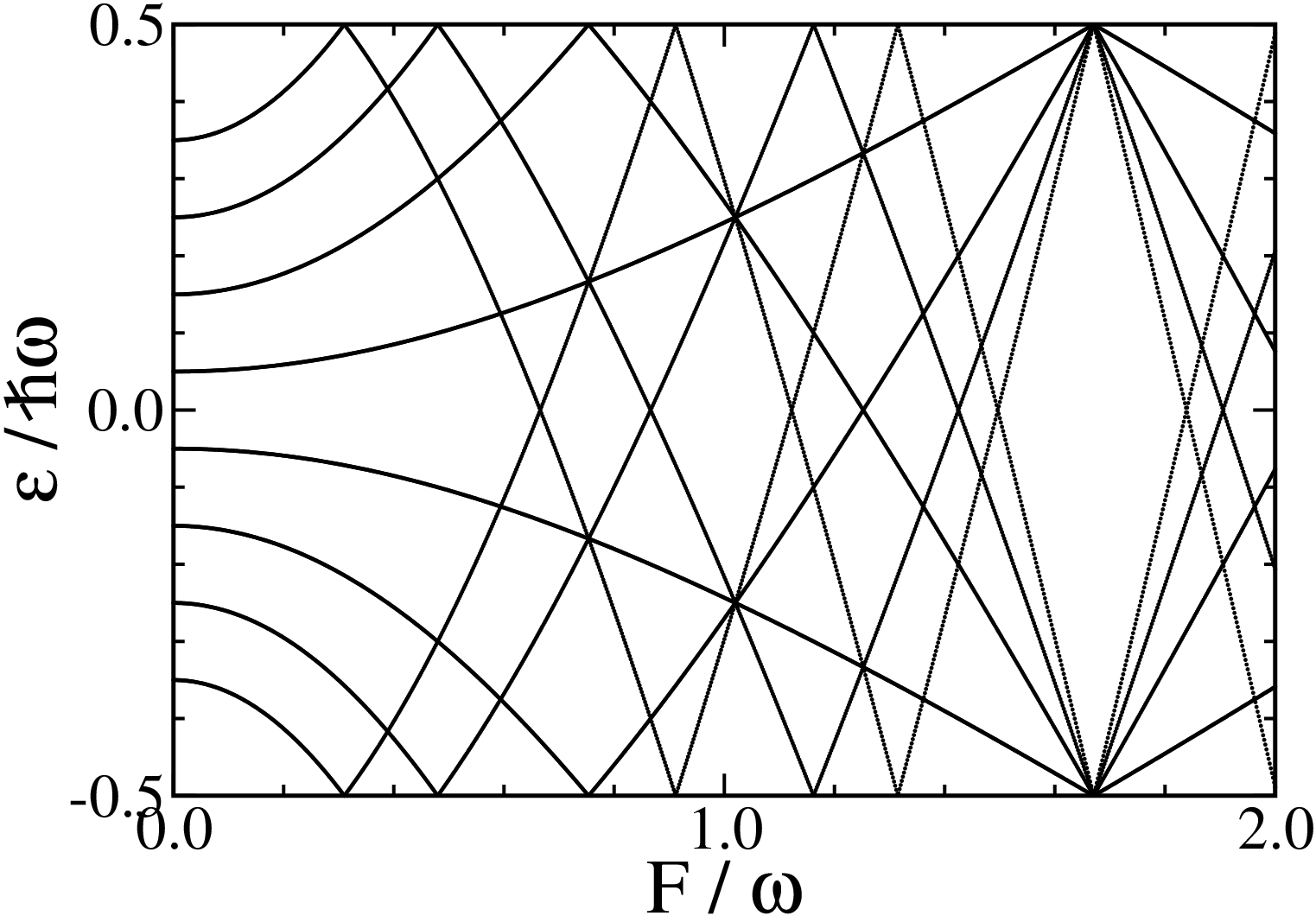}
\caption{One Brillouin zone of quasienergies for a spin $J = 7/2$ driven by a 
	left-circularly polarized high-frequency magnetic field according to 
	Eq.~(\ref{eq:LCP}), and exposed to a static magnetic field of scaled 
	strength $\omega_0/\omega = 0.1$. High degeneracies are found at 
	$F/\omega \approx 1.02$ ($\Omega/\omega = 3/2$) and
	$F/\omega \approx 1.67$ ($\Omega/\omega = 2$).}   
\label{F_5}
\end{figure}

\begin{figure}[t]
\centering
\includegraphics[width=0.9\linewidth]{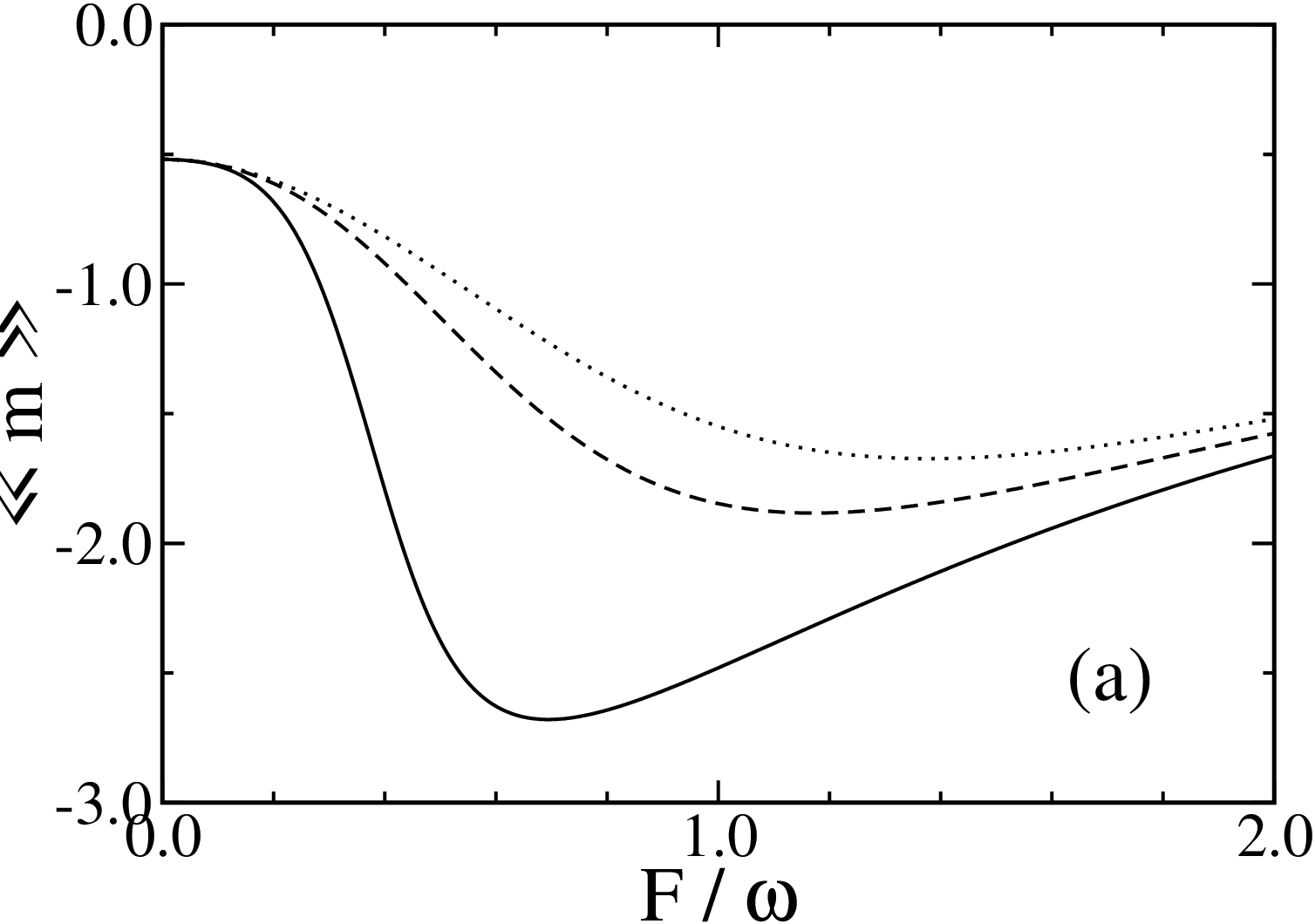}

\vspace{7ex}

\includegraphics[width=0.9\linewidth]{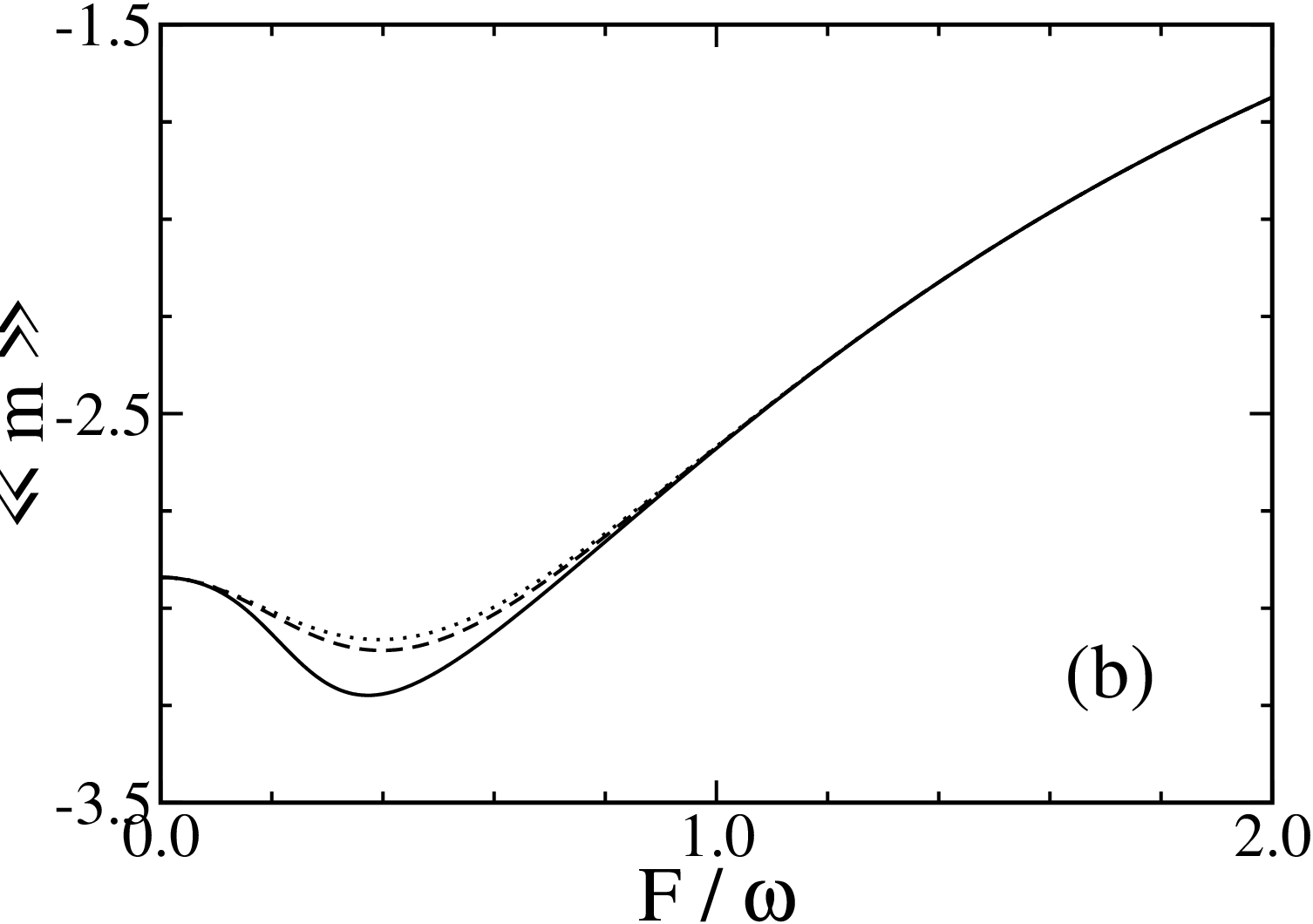}
\caption{As Fig.~\ref{F_2}, but for left-circularly polarized driving 
	described by Eq.~(\ref{eq:LCP}). Observe that the quasithermal
	magnetization is enhanced, so that the system is effectively cooled
	through the application of moderately strong fields.}  
\label{F_6}
\end{figure}

Further information on the interplay of the quasienergy spectrum and the 
quasithermal magnetization is obtained when replacing the right-circularly 
polarized driving field~(\ref{eq:CPF}) by a left-circularly polarized one,
as given by 
\begin{equation}
	H_{\rm osc}^{(l)}(t) = 
	\hbar F \big( S_x \cos(\omega t) - S_y \sin(\omega t) \big) \; .
\label{eq:LCP}	
\end{equation}
In order to elucidate the underlying symmetry properties, we consider the 
quasienergy operator for right-circular driving, that is, the operator 
\begin{equation}
	K^{(r)} = H_0 + H_{\rm osc}^{(r)}(t) - \ri\hbar\frac{\rd}{\rd t}
\label{eq:OKR}
\end{equation}
which determines the corresponding $2\pi/\omega$-periodic Floquet functions
$| u_m(t) \rangle$ and their quasienergies~$\varepsilon_m$ as solutions 
of the eigenvalue equation 
\begin{equation}
	K^{(r)} |u_m(t) \rangle = \varepsilon_m | u_m(t) \rangle \; ,	
\end{equation}
and subject this operator~(\ref{eq:OKR}) to the spatio-temporal parity 
operation
\begin{equation}
	{\mathcal PT} : \left\{ \begin{array}{rcl}
		\vec{S} & \to & -\vec{S} \\
		t	& \to & -t
	\end{array} \right.
\label{eq:PTS}
\end{equation}
simultaneously reversing the sign of the components $S_x$, $S_y$, $S_z$ of
the spin vector~$\vec{S}$, and that of the time coordinate~$t$. Under this
operation, one has 
\begin{equation}
	K^{(r)} \to -K^{(l)} \; ,
\end{equation}
where $K^{(l)}$ is the quasienergy operator for left-circular driving, being 
obtained from  $K^{(r)}$ through the replacement of $H_{\rm osc}^{(r)}(t)$ by 
its left-handed counterpart~(\ref{eq:LCP}). Now the operation~(\ref{eq:PTS})
can be separated into two consecutive steps: {\em (i)} a sign change of 
$S_z$ alone, as corresponding to a reversal of the direction of the static 
field~$B_0$, and {\em (ii)} the combined remaining sign changes of $S_x$, 
$S_y$, and~$t$, as corresponding to a transformation of $H_{\rm osc}^{(r)}(t)$
into $-H_{\rm osc}^{(l)}(t)$. Since the sign of $H_{\rm osc}(t)$ does not
affect the quasienergy spectrum in any case, and $-K^{(l)}$ is isospectral to 
$K^{(l)}$, this means that the quasienergy spectrum for {\em left\/}-circular 
driving with {\em positive\/}~$\omega_0$ is the same as that for 
{\em right\/}-circular driving with {\em negative\/}~$\omega_0$, and therefore
again is given by Eq.~(\ref{eq:QEC}). This is intuitively clear, since changing
the ``handedness'' of the circularly polarized drive should have the same
effect as a reversal of the direction of the orthogonally applied static
field. By analogous reasoning, the quasithermal magnetization for 
{\em left\/}-circular driving with {\em positive\/}~$\omega_0$ is the negative
of that for {\em right\/}-circular driving with {\em negative\/}~$\omega_0$.
But with $\omega \gg -\omega_0 > 0$, the canonical representatives of 
the quasienergies~(\ref{eq:QEC}) {\em repel\/} each other with increasing 
driving strength, as shown in Fig.~\ref{F_5}, so that there is no quasienergy 
collapse. As a consequence, the quasithermal magnetization induced by a
left-circularly polarized driving field is substantially different from that 
due to a right-circular drive, as exemplified by Fig.~\ref{F_6} for the same
conditions previously considered in Fig.~\ref{F_2}, except for the handedness 
of polarization. As expected, there is no sign change of the quasi\-thermal 
magnetization plotted in Fig.~\ref{F_6}, reflecting the absence of the 
principal resonance. However, for moderate driving amplitudes a noteworthy 
increase of the magnetization is achieved, reflecting ``Floquet-state cooling 
by driving''~\cite{DiermannHolthaus19}. This effect is particularly visible 
for $k_{\rm B}T/(\hbar\omega) = 1.0$ and the Gaussian density of states;
in that case the maximum quasi\-thermal magnetization exceeds the equilibrium
magnetization by a factor of more than five.

\section{An experimentum crucis}
\label{S_4}

Periodic thermodynamics of driven quantum systems, as envisioned by 
Kohn~\cite{Kohn01}, constitutes an area somewhere in between equilibrum 
thermodynamics on the one hand, and nonequilibrium thermodynamics of more 
general systems on the other. The fact that such periodically driven systems 
possess a basis of states --- the Floquet states --- which are occupied 
according to certain distributions $\{ p_n \}$ when being in contact with a 
heat bath is a feature they share with equilibrium systems. The fact that these
distributions are not universal, but do depend on the very details of the 
system-bath interaction, distinguishes them sharply from equilibrium systems, 
and can be exploited for deliberate quantum engineering. 

The present model study has identified consequences of periodic thermodynamics 
for ideal paramagnetic or diamagnetic spin systems in contact with a thermal 
environment, which are well amenable to accurate experimental investigation 
ever since Henry's seminal measurements~\cite{Henry52}. The ``quasithermal'' 
magnetization exhibited by such systems under the influence of a strong 
oscillating magnetic field shows at least four characteristic signatures:

\begin{itemize}
\item Different paramagnetic materials which respond in precisely the 
	same universal way to a static magnetic field in thermal 
	equilibrium~\cite{Brillouin27,Pathria11} may respond differently in 
	the presence of an additional strong oscillating field, depending on 
	how the spins interact with their environment.
\item In case of a circularly polarized driving field, quasi\-thermal 
	magnetization is strongly sensitive to the former's handedness.		
\item The quasithermal magnetization in the presence of periodic driving 
	can be even larger than the equilibrium thermal magnetization in 
	the absence of the drive, as if the system was effectively 
	cooled~\cite{DiermannHolthaus19}.
\item Even the sign of the quasithermal magnetization can be different from
	that of the equilibrium thermal magnetization~\cite{SchmidtEtAl19}, 
	reflecting a quasienergy collapse. 	  	 
\end{itemize}

Verification of these signatures constitutes an {\em experimentum crucis\/} 
for periodic thermodynamics: They are so elemental that they simply have to 
exist if the underlying theoretical framework is correct.  

Extensive experimental measurements of quasithermal magnetization will not only
be of conceptual merit, but also provide information not available otherwise,
because quasithermal magnetic response encodes the details of the mechanism 
by which the driven spins are interacting with their surroundings. While these
details necessarily have to remain hidden in measurements of ordinary thermal 
magnetism, due to the censorship imposed by the universality of equilibrium 
thermodynamics, the non-universality of periodic thermodynamics has the power
to lift this censorship, and to make them visible.

\begin{acknowledgments}
This work has been supported by the Deutsche For\-schungsgemeinschaft (DFG,
German Research Foundation) through Project No.~397122187. We wish to thank 
the members of the Research Unit FOR~2692 for many stimulating discussions. 
\end{acknowledgments}


\begin{thebibliography}{99}

\bibitem{Brillouin27}
	L. Brillouin,
	{\em Les moments de rotation et le magn\'etisme dans la
	m\'ecanique ondulatoire\/},
	J. Phys. Radium  {\bf 8}, 74 (1927).

\bibitem{Pathria11}
	For a modern exposition see, e.g.,
	R. K. Pathria and Paul D. Beale,
	{\em Statistical Mechanics\/}
	(Academic Press, New York; 3rd edition, 2011).

\bibitem{Henry52}
	W. E. Henry,
	{\em Spin paramagnetism of $\mbox{Cr}^{+++}$, $\mbox{Fe}^{+++}$, and 
	$\mbox{Gd}^{+++}$ at liquid Helium temperatures and in strong magnetic 
	fields\/},
	Phys. Rev. {\bf 88}, 559 (1952).
	
\bibitem{Balanda13}
	M. Ba{\l}anda,
	{\em AC susceptibility studies of phase transitions and magnetic
	relaxation: Conventional, molecular and low-dimensional magnets\/},
	Acta Phys. Pol. A {\bf 124}, 964 (2013).
	
\bibitem{KlikEtAl18}
	I. Klik, J. McHugh, R. W. Chantrell, and C.-R. Chang,
	{\em Debye formulas for a relaxing system with memory\/},		
	Sci. Rep.\ {\bf 8}, 3271 (2018).
	
\bibitem{ToppingBlundell19}
	C. V. Topping and S. J. Blundell,
	{\em A.C. susceptibility as a probe of low-frequency magnetic 
	dynamics\/},
	J. Phys.: Condens. Matter {\bf 31}, 013001 (2019).	 	

\bibitem{Kohn01}
	W. Kohn,
	{\em Periodic Thermodynamics\/},
	J. Stat. Phys. {\bf 103}, 417 (2001).
	
\bibitem{GrahamHuebner94}
	R. Graham and R. H\"ubner,
	{\em Generalized quasi-energies and Floquet states for a dissipative
	system\/},
	Ann. Phys. (New York) {\bf 234}, 300 (1994).
	
\bibitem{BreuerPetruccione97}
	H.-P. Breuer and F. Petruccione,
	{\em Dissipative quantum systems in strong laser fields: Stochastic
	wave-function method and Floquet theory\/},
	Phys. Rev. A {\bf 55}, 3101 (1997). 		
		
\bibitem{HoneEtAl09}
	D. W. Hone, R. Ketzmerick, and W. Kohn,
	{\em Statistical mechanics of Floquet systems: The pervasive problem
	of near degeneracies\/},
	Phys. Rev. E {\bf 79}, 051129 (2009).	
	
\bibitem{KetzmerickWustmann10}
	R. Ketzmerick and W. Wustmann,
	{\em Statistical mechanics of Floquet systems with regular and chaotic
	states\/},
	Phys. Rev. E {\bf 82}, 021114 (2010).
			
\bibitem{GasparinettiEtAl13}
	S. Gasparinetti, P. Solinas, S. Pugnetti, R. Fazio, and J. P. Pekola,
	{\em Environment-governed dynamics in driven quantum systems\/},
	Phys. Rev. Lett. {\bf 110}, 150403 (2013).

\bibitem{LangemeyerHolthaus14}
	M. Langemeyer and M. Holthaus,
	{\em Energy flow in periodic thermodynamics\/},
	Phys. Rev. E {\bf 89}, 012101 (2014).

\bibitem{BulnesCuetaraEtAl15}
	G. Bulnes Cuetara, A. Engel, and M. Esposito,
	{\em Stochastic thermodynamics of rapidly driven systems\/},
	New J. Phys. {\bf 17}, 055002 (2015).
	
\bibitem{ShiraiEtAl15}
	T. Shirai, T. Mori, and S. Miyashita,
	{\em Condition for emergence of the Floquet-Gibbs state in
	periodically driven open systems\/},
	Phys. Rev. E {\bf 91}, 030101(R) (2015).
	
\bibitem{RestrepoEtAl16}
	S. Restrepo, J. Cerrillo, V. M. Bastidas, D. G. Angelakis, and 
	T. Brandes,
	{\em Driven open quantum systems and Floquet stroboscopic dynamics\/},
	Phys. Rev. Lett. {\bf 117}, 250401 (2016).

\bibitem{HartmannEtAl17}	
	M. Hartmann, D. Poletti, M. Ivanchenko, S. Denisov, and P. H\"anggi, 	
	{\em Asymptotic Floquet states of open quantum systems: The role of 
	interaction\/},
	New J. Phys. {\bf 19}, 083011 (2017).
	
\bibitem{ZhangEtAl17} J. Zhang, P. W. Hess, A. Kyprianidis, P. Becker, A. Lee, 
	J. Smith, G. Pagano, I.-D. Potirniche, A. C. Potter, A. Vishwanath, 
	N. Y. Yao, and C. Monroe,
	{\em Observation of a discrete time crystal\/},
	Nature {\bf 543}, 217 (2017).
	
\bibitem{ChoiEtAl17} S. Choi, J. Choi, R. Landig, G. Kucsko, H. Zhou, J. Isoya,
	F. Jelezko, S. Onoda, H. Sumiya, V. Khemani, C. von Keyserlingk, 
	N. Y. Yao, E. Demler, and M. D. Lukin,
	{\em Observation of discrete time-crystalline order in a disordered 
	dipolar many-body system\/},
	Nature {\bf 543}, 221 (2017).	

\bibitem{Schmidt20}
	H.-J. Schmidt,
	{\em Periodic thermodynamics of a two spin Rabi model\/},
	J. Stat. Mech. (2020) 043204.
	
\bibitem{IkedaSato20}
	T. N. Ikeda and  M. Sato,
	{\em General description for non\-equilibrium steady states in 
	periodically driven dissipative quantum systems\/},
	arXiv:2003.02876 [cond-mat.stat-mech].	

\bibitem{BreuerEtAl00}
	H.-P. Breuer, W. Huber, and F. Petruccione,
	{\em Quasistationary distributions of dissipative nonlinear quantum
	oscillators in strong periodic driving fields\/},
	Phys. Rev. E {\bf 61}, 4883 (2000).	

\bibitem{DiermannEtAl19} 
	O. R. Diermann, H. Frerichs, and M. Holthaus,
	{\em Periodic thermodynamics of the parametrically driven harmonic
	oscillator\/},
	Phys. Rev. E {\bf 100}, 012102 (2019). 
	
\bibitem{DiermannHolthaus19}
	O. R. Diermann and M.~Holthaus,		
	{\em Floquet-state cooling\/}, 
	Sci. Rep.\ {\bf 9}, 17614 (2019).  
				
\bibitem{SchmidtEtAl19}
	H.-J. Schmidt, J. Schnack, and M. Holthaus,		
      	{\em Periodic thermodynamics of the Rabi model with circular 
	polarization for arbitrary spin quantum numbers\/},
	Phys. Rev. E {\bf 100}, 042141 (2019).	
		
\bibitem{BreuerPetruccione02}
	H.-P. Breuer and F. Petruccione,
	{\em The Theory of Open Quantum Systems\/}
	(Oxford University Press, Oxford, 2002).
	
\bibitem{Holthaus92}
	M. Holthaus,
	{\em The quantum theory of an ideal superlattice responding to
	far-infrared laser radiation\/},
	Z. Phys. B {\bf 89}, 251 (1992).	

\end{thebibliography}
\end{document}